\RequirePackage{fix-cm}
\documentclass[twocolumn]{svjour3}          % twocolumn
\smartqed  % flush right qed marks, e.g. at end of proof
\usepackage{graphicx}

\graphicspath{{Bilder/}} 
\usepackage{amsmath}
\usepackage{amssymb}
\usepackage{braket}
\usepackage{subfigure}
\usepackage{url}
\usepackage{widetext}
\usepackage{color}
\let \Re \relax \DeclareMathOperator{\Re}{Re}
\let \Im \relax \DeclareMathOperator{\Im}{Im}
\newcommand{\ii}{\mathrm{i}}
 \journalname{EPJB}

\begin{document}

\title{Real and imaginary edge states in stacked Floquet honeycomb lattices}
\subtitle{}


\author{Alexander Fritzsche \and Bastian H{\"o}ckendorf \and Andreas Alvermann \and Holger Fehske}


\institute{Alexander Fritzsche \at
              Institut f{\"u}r Theoretische Physik und Astrophysik, Julius-Maximilians-Universit{\"a}t W{\"u}rzburg, Am Hubland, 97074 W{\"u}rzburg, Germany\\
              \email{alexander.fritzsche@physik.uni-wuerzburg.de}           %  \\
           \and
           Bastian H{\"o}ckendorf \at
            Institut f{\"u}r Physik, Universit{\"a}t Greifswald, Felix-Hausdorff-Str. 6, 17489 Greifswald, Germany
           \and 
           Andreas Alvermann \at 
           Institut f{\"u}r Physik, Universit{\"a}t Greifswald, Felix-Hausdorff-Str. 6, 17489 Greifswald, Germany
           \and
           Holger Fehske \at
            Institut f{\"u}r Physik, Universit{\"a}t Greifswald, Felix-Hausdorff-Str. 6, 17489 Greifswald, Germany\\
}

\date{Received: date / Accepted: date}

\maketitle

\begin{abstract}
We present a non-Hermitian Floquet model with topological edge states in real and imaginary band gaps. 
The model utilizes two stacked honeycomb lattices which can be related via four different types of non-Hermitian time-reversal symmetry. Implementing the correct time-reversal symmetry provides us with either two counterpropagating edge states in a real gap, or a single edge state in an imaginary gap. The counterpropagating edge states allow for either helical or chiral transport along the lattice perimeter. In stark contrast, we find that the edge state in the imaginary gap does not propagate. 
Instead, it remains spatially localized while its amplitude continuously increases.
Our model is well-suited for realizing these edge states in photonic waveguide lattices.
\end{abstract}

\section{Introduction}
\label{intro}

After their discovery in 1980 \cite{Klitzing} topological states of matter have been in the focus of condensed matter research for the past decades and have led to fundamental insights regarding the interplay between bulk topology and edge transport \cite{molenkamp,qizhangTI,reviewTI}.  
Unidirectional transport emerges at an edge via chiral edge states if the bulk is topologically non-trivial.
In combination with time-reversal symmetry, topology even allows for bidirectional helical transport via counterpropagating edge states \cite{Schnyder,pertableherm,Kanemele,Fukane,Lababidi,lukasbastian}.  
With the discovery of anomalous Floquet topological insulators~\cite{Kitagawa,Rudner1,Rudner2,Mukherjee2017,Maczewsky,BastianW3}, it was found that time-periodicity leads to another unique interplay between bulk topology and edge transport in the form of quantized charge pumping~\cite{PhysRevX.6.021013}.

It was recognized only recently that non-Hermiticity extends this picture even further, both for static~\cite{BenderNH,NHEsaki,NHRonny,NHYao,Lee} and Floquet systems~\cite{StateEngineering,fedorova2019topological,hckendorf2019nonhermitian,hckendorf2020cutting}. In addition to the familiar topological phases with real band gaps, new non-Hermitian topological phases emerge~\cite{Kawabata1,Kawabata2,pertableUSA} which arise from imaginary and point gaps in the complex-valued spectrum. While the topological phases arising from point gaps have been extensively investigated theoretically~\cite{NHRonny,Kawabata1} and experimentally~\cite{circuits,Weidemann311}, studies on edge states in imaginary gaps and their transport properties are still rare. 

In this paper, we present a non-Hermitian Floquet model which possesses edge states in both real and imaginary gaps.
The model consists of two layers of honeycomb lattices stacked on top of each other.  Two inverse copies of an anomalous Floquet topological insulator are implemented on the two layers. 
This arrangement is inherently time-reversal symmetric. In combination with complex on-site potentials, which introduce non-Hermiticity, four distinct types of non-Hermitian time-reversal symmetry can occur within the model. The first two symmetries enforce counterpropagating edge states in real gaps, the third one is incompatible with topological edge states, and the fourth one allows for a single edge state in an imaginary gap. 

We discuss how our model could be implemented in photonic waveguide lattices where the real-space propagation of edge states is directly observable. In contrast to the counterpropagating edge states in real gaps, we find that the edge state in the imaginary gap remains localized during time evolution while its amplitude continuously increases. This amplification  is protected by time-reversal symmetry. In this way, the symmetry and its realization in a double-layer honeycomb lattice is essential for this edge state phenomenon.

This paper is organized as follows. 
In Sec.~\ref{sec:model}, we introduce the stacked honeycomb model. In Sec.~\ref{sec:TRS}, we implement the four time-reversal symmetry types and determine the resulting constraints on the model parameters. We provide specific parameter values for the different cases and demonstrate the existence of edge states in real and imaginary gaps. The three-dimensional geometry of the two honeycomb layers can not be implemented in photonic waveguide lattices, which is why we map the two honeycomb layers onto a square lattice in Sec.~\ref{sec:prop}.  For this lattice configuration, the propagation of the edge states is investigated.  We conclude in Sec.~\ref{sec:conclusion}.

\section{Stacked honeycomb model}
\label{sec:model}

The basis of our studies is an extension of the Floquet driving protocol presented in Ref.~\cite{Kitagawa}, which consists of three time steps that cycle through the three nearest-neighbor couplings on a honeycomb lattice. In each step, two of the three couplings are set to zero while the third coupling has a constant non-zero value. Depending on the coupling strength, the system is either in a trivial phase or in an anomalous Floquet topological phase with a single edge state in a real gap~\cite{Kitagawa}.

We would like to point out that this driving protocol can not be implemented in graphene~\cite{reviewgraphene}. Periodically switching the nearest-neighbor couplings on and off is currently not possible in a real solid state system. In addition, next-nearest-neighbor couplings, spin-orbit coupling, and interactions have to be included in a graphene model which are not considered here. 

\begin{figure}
\centering
\includegraphics[scale=.5]{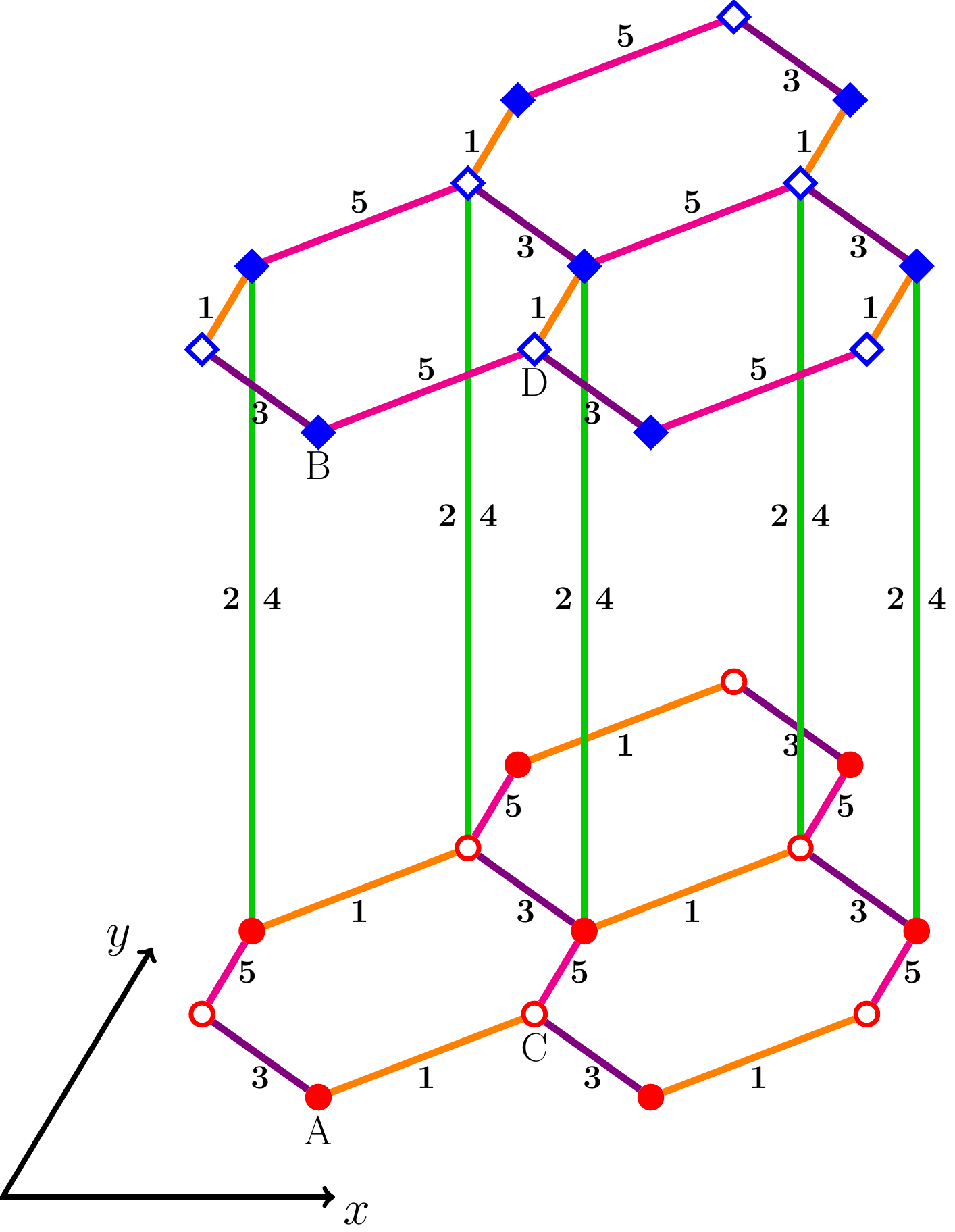}
\caption{Double-layer Floquet honeycomb model~\eqref{eq:model}. The two layers are stacked on top of each other. The lattice is periodic along the $x$-axis and $y$-axis. The model consists of a cyclically repeated sequence of five time steps.  In steps $1$ (orange), $3$ (violet), and $5$ (magenta) intralayer couplings occur while interlayer couplings occur in steps $2$ and $4$ (both green).  For clarity, only a few of the interlayer couplings are shown here. The unit cell is given by the four sites labeled with $A$, $B$, $C$, and $D$. }
\label{model}
\end{figure}

In our model, we add a second honeycomb layer on which an inverse copy of the driving protocol is implemented, similiar to the procedure in Refs.~\cite{lukasbastian,Bastian2}. 
The two layers (indicated by red circles and blue diamonds in Fig.~\ref{model}) are then coupled in two additional time steps where all intralayer couplings are set to zero. In total, one period $T$ in our model cycles through five time steps of equal length $T/5$ with pairwise coupling in each step. This precise control over the couplings becomes possible in photonic waveguides~\cite{lukasbastian,Mukherjee2017,Maczewsky}. 

The position of the two honeycomb layers relative to each other is irrelevant for the theoretical description of the  interlayer coupling. For now, we utilize the stacked geometry in Fig.~\ref{model} to graphically separate the two layers into a ``bottom'' and a ``top'' layer. In Sec.~\ref{sec:prop}, we switch to a planar geometry, which is more relevant for experiments in photonic systems.

The time-periodic Bloch Hamiltonian can be expressed in standard bra-ket notation as
 \begin{equation}
\begin{aligned}
H(\vec k,t)=&\sum_{l=1}^3 \Big[J_l(t) e^{\ii \vec k \cdot \boldsymbol \delta_l}| \vec k,A \rangle \langle \vec k,C |+\mathrm{H.c.}\Big]
\\
+&\sum_{l=1}^3\Big[J_l(T-t) e^{\ii \vec k \cdot \boldsymbol \delta_l} |\vec k,B \rangle \langle \vec k,D | + \mathrm{H.c.}\Big]
\\
+&J'(t)\Big(|\vec k,A \rangle \langle \vec k,B|+|\vec k,C \rangle \langle \vec k,D|\Big) + \mathrm{H.c.}\\
+& \sum_{s\in \{A,B,C,D\}} \Delta_s |\vec k, s\rangle \langle \vec k, s|  \; .
\end{aligned}
\label{eq:model}
\end{equation}
The first line of Eq.~\eqref{eq:model}  includes the three nearest-neighbor couplings $J_l(t)$ between sites $A$, $C$ on the bottom layer along the directions $\boldsymbol \delta_1=(\sqrt3/2,1/2)$, $\boldsymbol \delta_2=(-\sqrt3/2,1/2)$, and $\boldsymbol \delta_3=(0,-1)$. In steps $1$, $3$, and $5$, exactly one of these three couplings is set to the constant real value $J_l(t)=J$ while the other two couplings are set to zero (see Fig.~\ref{model}). In the second line, the same couplings are used for the top layer with sites $B$, $D$, but in inverse time order.
 In the third line, the interlayer coupling $J'(t)$ appears which is non-zero only in steps $2$ and $4$. We set $J'(t)=J$ in step $2$ and $J'(t)=\alpha J$ in step $4$ with the sign $\alpha=\pm 1$ as a free parameter.

For vanishing on-site potentials $\Delta_s=0$, the propagator 
\begin{equation}
U_J=\begin{pmatrix} \cos (JT/5) & -\ii \sin (JT/5) \\ -\ii \sin (JT/5) & \cos (JT/5) \end{pmatrix}
\end{equation}
of two coupled sites is a periodic function of $J$ for each time step of length $T/5$.
The time steps in Eq.~\eqref{eq:model} are arranged such that edge states on different layers propagate in opposite directions at perfect coupling ($J=5\pi/(2T), \Delta_s=0$), where a full amplitude transfer occurs between two coupled sites in each step.  An edge state on the bottom layer propagates in counter-clockwise direction along the lattice perimeter while an edge state on the top layer moves clockwise. The bulk states are completely localized on both layers.

The two counterpropagating edge states appear as chiral modes in the real band gaps of the quasienergy spectrum $\{\varepsilon(\vec k)\}$, which is obtained from the eigenvalues $\{\mathrm{e}^{- \ii \varepsilon(\vec{k})T}\}$ of the Floquet-Bloch propagator 
\begin{equation}
U(\vec k,T)=\mathcal{T}\mathrm{exp}\Big(-\ii \int_0^T H(\vec k,t')\mathrm{d}t'\Big)
\end{equation}
 ($\mathcal{T}$ denotes time-ordering, and we set $\hbar \equiv 1$). Due to the $2\pi/T$-periodicity of the eigenvalues, we restrict the real part of the quasienergy spectrum to the quasienergy Brillouin zone $-\pi/T\leq \Re \varepsilon(\vec{k})\leq \pi/T $.

The complex on-site potentials $\Delta_s$ in the fourth line of Eq.~\eqref{eq:model} introduce non-Hermiticity which leads to complex quasienergies. In addition to the aforementioned counterpropagating edge states in real gaps, edge states in imaginary gaps can be obtained for non-zero on-site potentials. Which of the two cases occurs for a specific set of parameters is linked to the fundamental symmetries of our model.

\section{Floquet bands, edge states and time-reversal symmetry}
\label{sec:TRS}

\begin{table}
\caption{Parameter sets for the stacked Floquet honeycomb model with $\mathrm{TRS}^*$ or $\mathrm{TRS}^T$. The sign $\alpha$ determines if the symmetry is bosonic or fermionic.}
\begin{tabular}{lll}
\hline\noalign{\smallskip}
fermionic $\mathrm{TRS}^*$ & fermionic $\mathrm{TRS}^T$ \\
\noalign{\smallskip}\hline\noalign{\smallskip}
$~\;\; \alpha=-1$&$~\;\; \alpha=-1$\\
$\Delta_\mathrm{A}=-\Delta_\mathrm{C}=\delta+i\gamma$ & $\Delta_\mathrm{A}=-\Delta_\mathrm{C}=\delta+i\gamma$ \\
$\Delta_\mathrm{B}=-\Delta_\mathrm{D}=\delta-i\gamma$ & $\Delta_\mathrm{B}=-\Delta_\mathrm{D}=\delta+i\gamma$ \\
\noalign{\smallskip}\hline\noalign{\smallskip}
$J=\frac{5\pi}{2T}\, , \, \delta=\frac{3}{4T}\, , \, \gamma=\frac{5}{2T}$& $J=\frac{5\pi}{2T}\, , \, \delta=\frac{3}{4T}\, , \, \gamma=\frac{5}{2T}$\\
\noalign{\smallskip}\hline\noalign{\smallskip}
bosonic $\mathrm{TRS}^*$ & bosonic $\mathrm{TRS}^T$   \\
\noalign{\smallskip}\hline\noalign{\smallskip}
$~\;\; \alpha=1$&$~\;\; \alpha=1$\\
$\Delta_\mathrm{A}=-\Delta_\mathrm{C}=\delta+i\gamma$ & $\Delta_\mathrm{A}=-\Delta_\mathrm{C}=\delta+i\gamma$ \\
$\Delta_\mathrm{B}=-\Delta_\mathrm{D}=\delta-i\gamma$ & $\Delta_\mathrm{B}=-\Delta_\mathrm{D}=\delta+i\gamma$ \\
\noalign{\smallskip}\hline\noalign{\smallskip}
$J=\frac{5\pi}{4T}\, , \, \delta=\frac{1}{2T}\, , \,\gamma=\frac{3}{2T}$ & $J=\frac{5\pi}{2.2T}\, , \, \delta=\frac{1}{2T}\, , \, \gamma=\frac{2}{T}$\\
\noalign{\smallskip}\hline
\label{parametersnh}
\end{tabular}
\end{table}

In a Hermitian system, the symmetry relation for time-reversal symmetry (TRS) is
\begin{equation}
H(-\vec{k},T-t)=\Theta H(\vec{k},t)\Theta^{-1}
\label{symmrel}
\end{equation}
with an anti-unitary operator $\Theta$ for which $\Theta^2=\pm 1$~\cite{pertableherm}. The anti-unitary operator $\Theta= \mathcal K \theta$ can be written as the product of the complex conjugation operator $\mathcal K$ and a unitary operator $\theta$ which satisfies $\theta^*\theta=\pm 1$. 
Per construction, Eq.~\eqref{model} with zero on-site potentials $\Delta_s =0$ is time-reversal symmetric.
Here, the operator $\theta$ exchanges the two honeycomb layers such that site $A$ ($C$) is mapped onto site $B$ ($D$) and vice versa.  For fermionic (bosonic) TRS with $\theta^*\theta=-1$ ($\theta^*\theta=1$) the sign of the interlayer coupling in step $4$ is set to $\alpha=-1$ ($\alpha=1$).

 \begin{figure}
 \hspace*{\fill}
\includegraphics[scale=1.2]{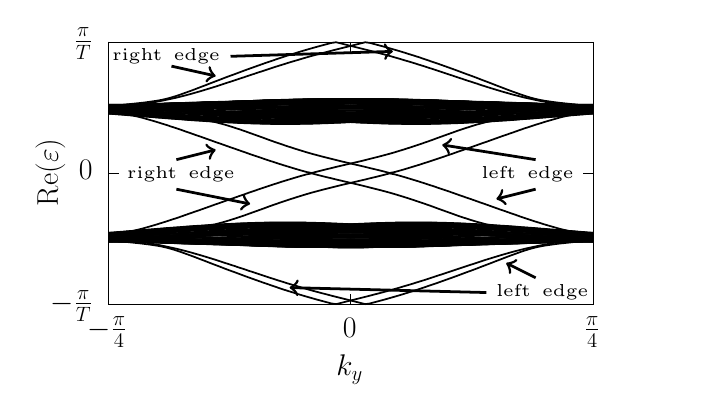} 
\hspace*{\fill}
\\
 \hspace*{\fill}
\includegraphics[scale=1.3]{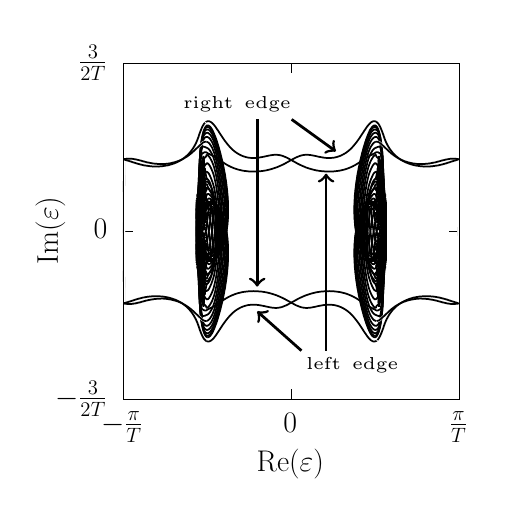}
 \hspace*{\fill}
\caption{Floquet bands and edge state dispersions for fermionic $\mathrm{TRS}^*$ with the parameters from Tab.~\ref{parametersnh}. Shown is the real part of the quasienergy dispersion as a function of momentum $k_y$ (top panel) and the imaginary part as a function of the real part (bottom panel).  We observe two counterpropagating edge states in the real gaps at $\Re \varepsilon= 0, \pi/T$ with opposite imaginary part. Here, and in Figs.~\ref{nhtrsdag}-\ref{imagstates}, we show the bands and edge states on a semi-infinite strip along the $y$-axis with armchair boundaries on both layers of the stacked honeycomb lattice. The arrows indicate which edge states belong to the left and right edge of the strip.}
\label{nhtrs}
\end{figure}

In the non-Hermitian case $H^*(\vec k,t)\ne H^T(\vec k,t)$ with complex on-site potentials, Eq.~\eqref{symmrel} splits up into two independent symmetry relations~\cite{StateEngineering,Kawabata2,pertableUSA}
\begin{subequations}
\label{TRS_NH}
\begin{align}
\label{TRS_NH_1}
\mathrm{TRS}^*: \quad H(-\vec k,T-t)&= \theta H^*(\vec k,t) \theta^{-1}  \;,\\
\label{TRS_NH_2}
\mathrm{TRS}^T: \quad H(-\vec k,T-t)&= \theta H^T(\vec k,t) \theta^{-1} \; .
\end{align}
\end{subequations}
Here, $(\cdot)^*$ denotes complex conjugation and $(\cdot)^T$ transposition. The two symmetry relations enforce the constraints
\begin{subequations}
\label{TRSdispersion}
\begin{align}
 \mathrm{TRS}^{*,T}: \quad \Re \, \{ \varepsilon(\vec k) \} & = \phantom{-} \Re \, \{ \varepsilon(-\vec k) \} \;, \label{TRSdispersionReal} \\
\mathrm{TRS}^*: \quad \Im \, \{ \varepsilon(\vec k) \}  &= - \Im \, \{ \varepsilon(- \vec k)  \}   \;, \label{TRSdispersionImC} \\
\mathrm{TRS}^T: \quad  \Im  \, \{\varepsilon(\vec k) \}  &= \phantom{-} \Im  \, \{  \varepsilon(- \vec k) \} 
\label{TRSdispersionImT}
\end{align}
\end{subequations}
upon the real and imaginary part of the quasienergy spectrum $\{ \varepsilon(\vec k) \}$. 
Note that the constraints enforced upon the real and imaginary part differ for $\mathrm{TRS}^*$. In that case, different topological phases can emerge in real and imaginary gaps. While Eq.~\eqref{TRSdispersionReal} implies that edge states appear in pairs with opposite chirality in real gaps, Eq.~\eqref{TRSdispersionImC} allows for individual edge states in imaginary gaps.

\begin{figure}
 \hspace*{\fill}
\includegraphics[scale=1.2]{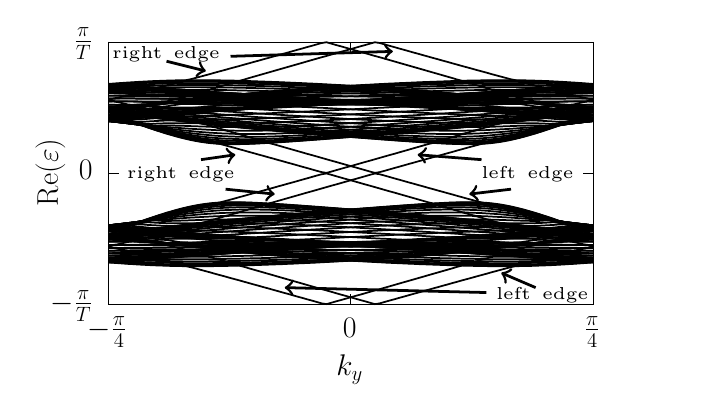}
 \hspace*{\fill}
 \\
  \hspace*{\fill}
\includegraphics[scale=1.3]{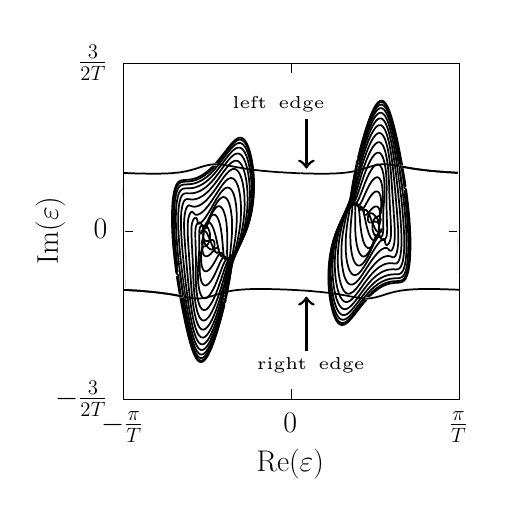}
 \hspace*{\fill}
\caption{Same as Fig.~\ref{nhtrs}, now for fermionic $\mathrm{TRS}^T$ with the parameters from Tab.~\ref{parametersnh}. We again observe two counterpropagating edge states in the real gaps at $\Re \varepsilon= 0, \pi/T$, but they now have the same imaginary part.}
\label{nhtrsdag}
\end{figure}

The symmetry relations~\eqref{TRS_NH} are fulfilled if we set $\Delta_A=\Delta_B^*$, $\Delta_C=\Delta_D^*$ for $\mathrm{TRS^*}$ and $\Delta_A=\Delta_B$, $\Delta_C=\Delta_D$ for  $\mathrm{TRS}^T$.  Combined with the possibility to switch between a bosonic and fermionic symmetry via the parameter $\alpha$, we get four distinct TRS types that can be realized in the present model.  
In the following, we will use the parameter sets in Tab.~\ref{parametersnh} to implement the four symmetry types and explore the ensuing topological phases.
It should be noted that in an experiment, fine-tuning to these parameter values is not required. As long as the relevant band gap does not close, one may continuously deform the parameters as desired.  However, the deformed parameters still have to satisfy the TRS relations in Eq.~\eqref{TRS_NH}.

In Figs.~\ref{nhtrs} and~\ref{nhtrsdag}, we show the Floquet bands and the edge state dispersions for fermionic $\mathrm{TRS}^*$ and fermionic $\mathrm{TRS}^T$, respectively. In both cases, we observe two counterpropagating edge states per edge which cross at momentum $k_y=0$ in the real gaps at $\Re \varepsilon= 0, \pi/T$. For $\mathrm{TRS^*}$, the two edge states are separated by their imaginary part at the crossing, while for $\mathrm{TRS^T}$, they cross at the same imaginary part. In that case, the crossing is protected by Kramers degeneracy. The counterpropagating edge states indicate a $\mathbb Z_2$ topological phase for the real gaps, which is in agreement with the symmetry classification in Ref.~\cite{Kawabata2}. Since the counterpropagating edge states appear in all real gaps of the quasienergy spectrum, the $\mathbb Z_2$ phase is anomalous~\cite{Rudner2}.

\begin{figure}
\hspace*{\fill}
\includegraphics[scale=1.2]{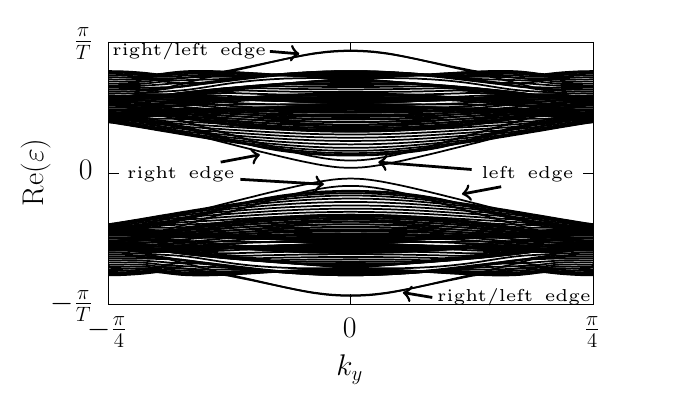}
\hspace*{\fill}
\\
\hspace*{\fill}
\includegraphics[scale=1.3]{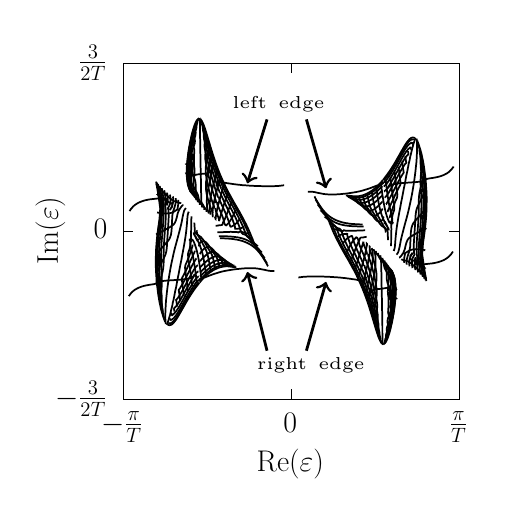}
\hspace*{\fill}
\caption{Same as Figs.~\ref{nhtrs},~\ref{nhtrsdag}, now for bosonic $\mathrm{TRS}^T$ with the parameters from Tab.~\ref{parametersnh}. The trivial edge states do not traverse the real gaps at $\Re \varepsilon= 0, \pi/T$.}
\label{nhbostrs}
\end{figure}

For the bosonic symmetries, the crossing at $k_y=0$ is not protected by Kramers degeneracy and the counterpropagating edge states can cancel, resulting in trivial edge states. We demonstrate this for bosonic $\mathrm{TRS}^T$ in Fig.~\ref{nhbostrs}. The trivial edge states do not traverse the real gaps at $\Re \varepsilon= 0, \pi/T$. Through appropriate symmetry-preserving parameter variations, the edge states could be continuously deformed to merge with the Floquet bands. These results also hold for bosonic $\mathrm{TRS}^*$.

\begin{figure}
 \hspace*{\fill}
\includegraphics[scale=1.2]{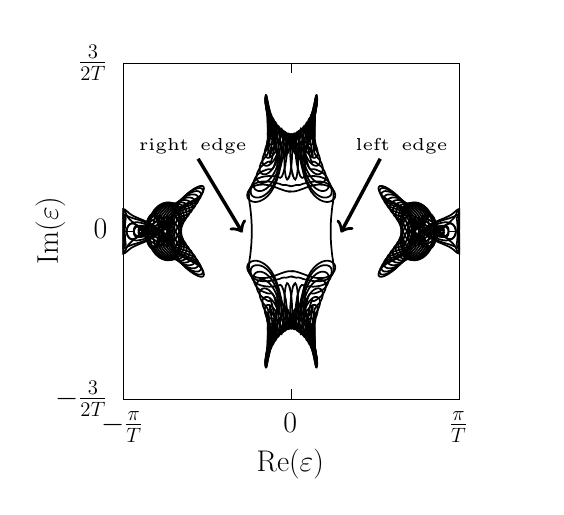}
 \hspace*{\fill}
 \\
  \hspace*{\fill}
\includegraphics[scale=1.3]{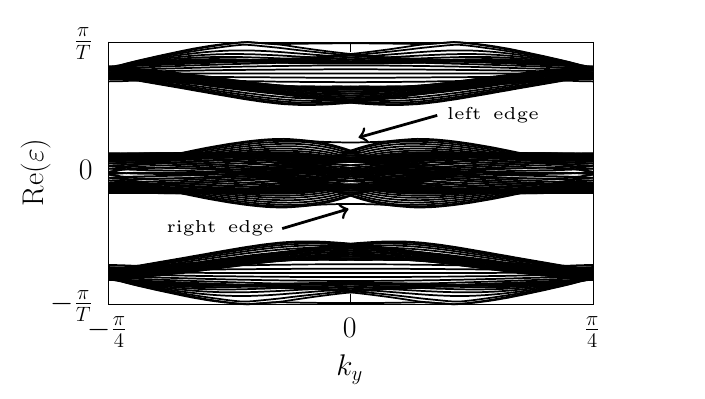}
 \hspace*{\fill}
 \\
  \hspace*{\fill}
\includegraphics[scale=1.3]{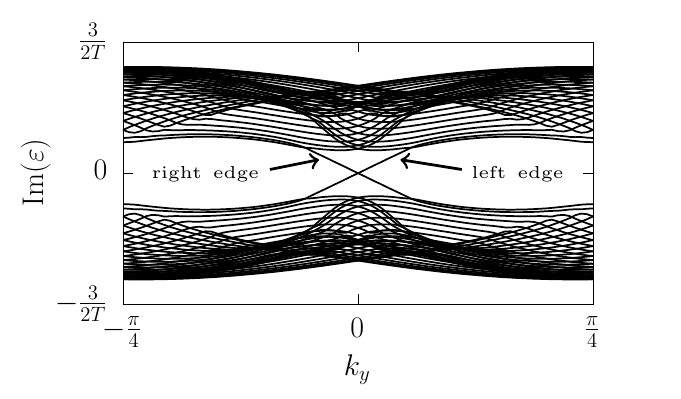}
 \hspace*{\fill}
\caption{Same as Figs.~\ref{nhtrs}-\ref{nhbostrs}, now for bosonic $\mathrm{TRS}^*$ with the parameters from Tab.~\ref{parametersnh}. Here, we include the imaginary part of the quasienergy dispersion as a function of $k_y$ (bottom panel). An edge state traverses the imaginary gap at $\Im \varepsilon=0$.}
\label{imagstates}
\end{figure}

In imaginary gaps, edge states become possible for bosonic $\mathrm{TRS}^*$. In Fig.~\ref{imagstates}, we observe a single edge state which traverses the imaginary gap at $\Im \varepsilon=0$. This unpaired edge state is pinned at momentum $k_y=0$ by the quasienergy relation~\eqref{TRSdispersionImC}. Note that only the imaginary part of the edge state dispersion is chiral. The real part is flat.  The edge state indicates a $\mathbb Z$ topological phase for the imaginary gap, which also agrees with the symmetry classification in Ref.~\cite{Kawabata2}. 

Unfortunately, the imaginary edge state is attached to a band which resides above the imaginary gap at $\Im \varepsilon=0$. This band necessarily has a larger imaginary part than the edge state and thus will dominate on large time scales, making the observation of the imaginary edge state challenging in experiments.

\section{Edge state propagation in a square lattice}
\label{sec:prop}

\begin{figure}
\hspace*{\fill}
\includegraphics[width=0.9\linewidth]{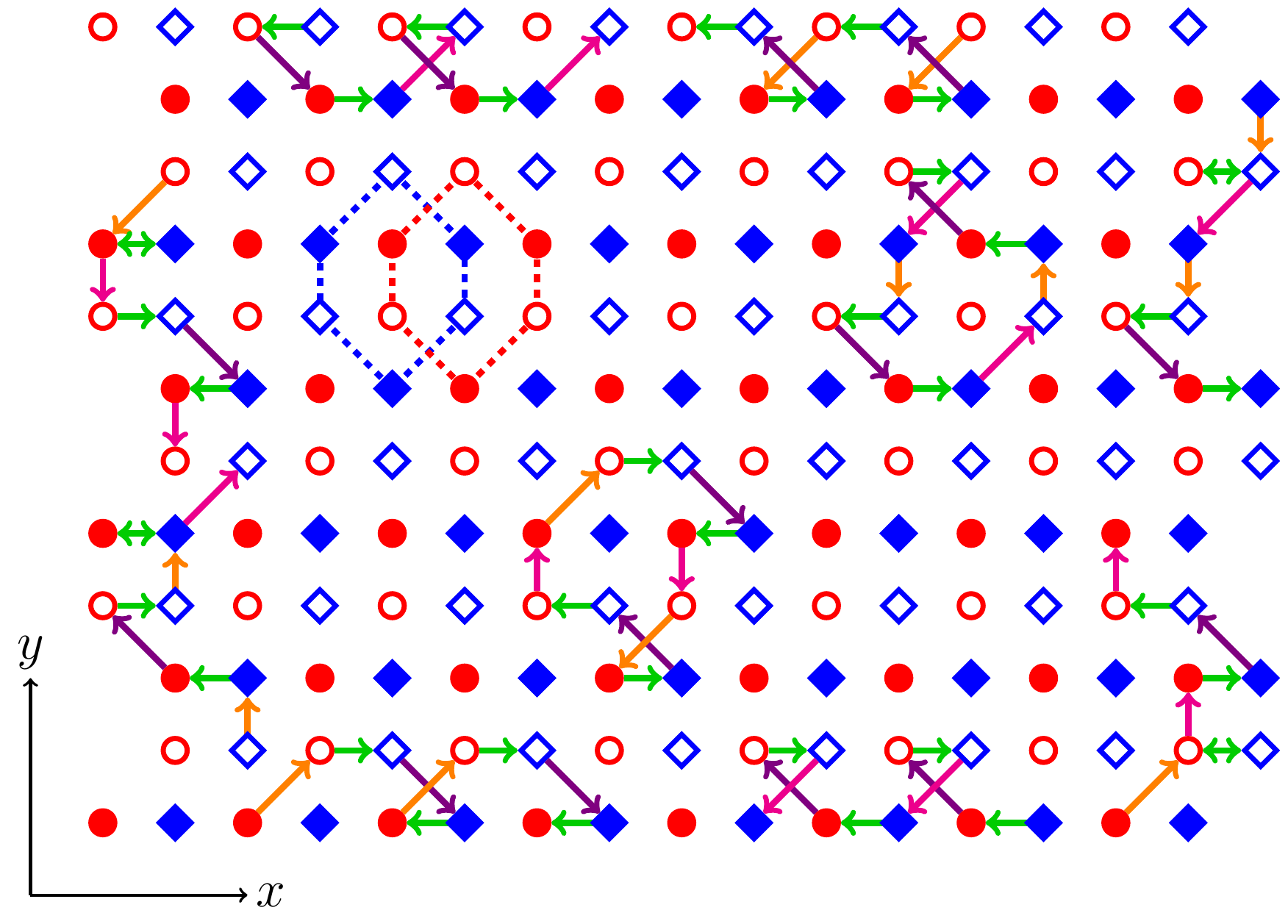}
\hspace*{\fill}
\caption{Patterns of motion at perfect coupling on a finite square lattice with eight unit cells in the $x$-direction and three unit cells in the $y$-direction. The edges parallel to the $x$-axis ($y$-axis) correspond to zigzag (armchair) edges on both layers of the stacked honeycomb geometry in Fig.~\ref{model}. The honeycomb structure of the red and blue sublattices are indicated by the dashed lines.
We show the trajectories for two full cycles of the time-periodic driving. Bulk states are localized while edge states propagate.} 
\label{trajec}
\end{figure}

Photonic waveguide lattices are well-suited for the realization of two-dimensional Floquet systems~\cite{lukasbastian,Mukherjee2017,Maczewsky}. Since the third spatial coordinate represents the time axis in waveguide lattices, the stacked geometry in Fig.~\ref{model} is inconvenient. Therefore, we now switch to a planar geometry where the two honeycomb layers are mapped onto a square lattice (see the dashed lines in Fig.~\ref{trajec}). In this way, our model  can be readily implemented in photonic lattices. The couplings $J(t)$, $J'(t)$ and complex on-site potentials $\Delta_s$ are then realized via spatially periodic modulation of the interwaveguide distance~\cite{Szameit} and manipulation of the waveguide losses~\cite{photonics3}, respectively. 

\begin{figure}
\centering
\includegraphics[width=1\linewidth]{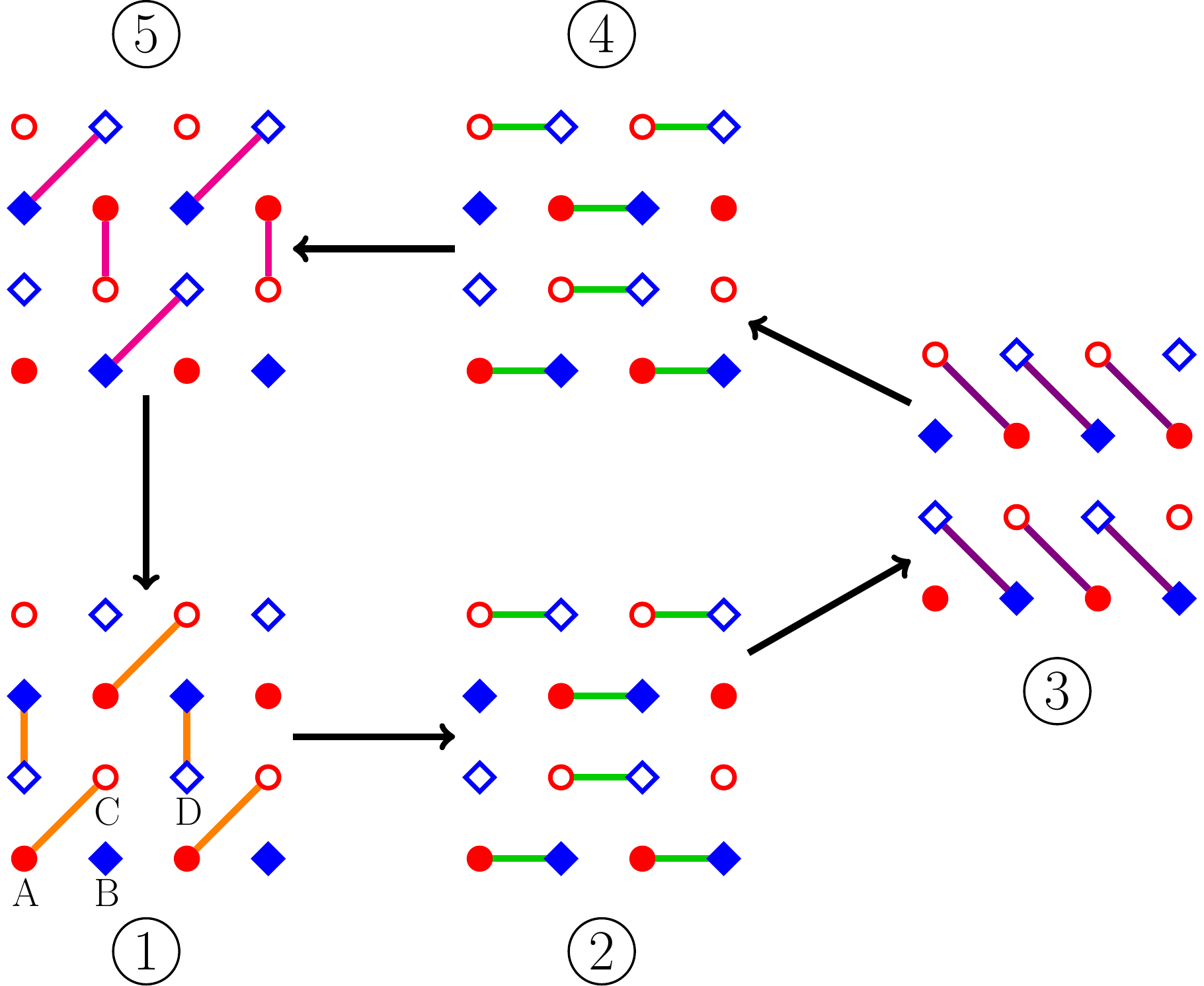}
\label{protocol2d}
\caption{The five time steps of Eq.~\eqref{eq:model} adapted for a two-dimensional square lattice. The red circles and blue diamonds correspond to the bottom and top honeycomb layer in Fig.~\ref{model}, respectively. The unit cell is given by the four sites labeled with $A$, $B$, $C$, and $D$.  }
\label{Fig:2d}
\end{figure}

In the square lattice, the two sites $A$, $C$ of the bottom honeycomb layer are located at the positions $\vec r_A=(i+2j,2i)$, $\vec r_C=\vec r_A+(1,1)$ with $i,j \in \mathbb Z$. The two sites $B, D$ of the top layer are located at $\vec r_B=\vec r_A+(1,0)$, $\vec r_D=\vec r_C+(1,0)$. The positions are chosen such that in each of the five time steps couplings only occur between neighboring sites (see Fig.~\ref{Fig:2d}). 

\begin{figure*}
\begin{minipage}[l]{0.5\textwidth}
\includegraphics[scale=.285]{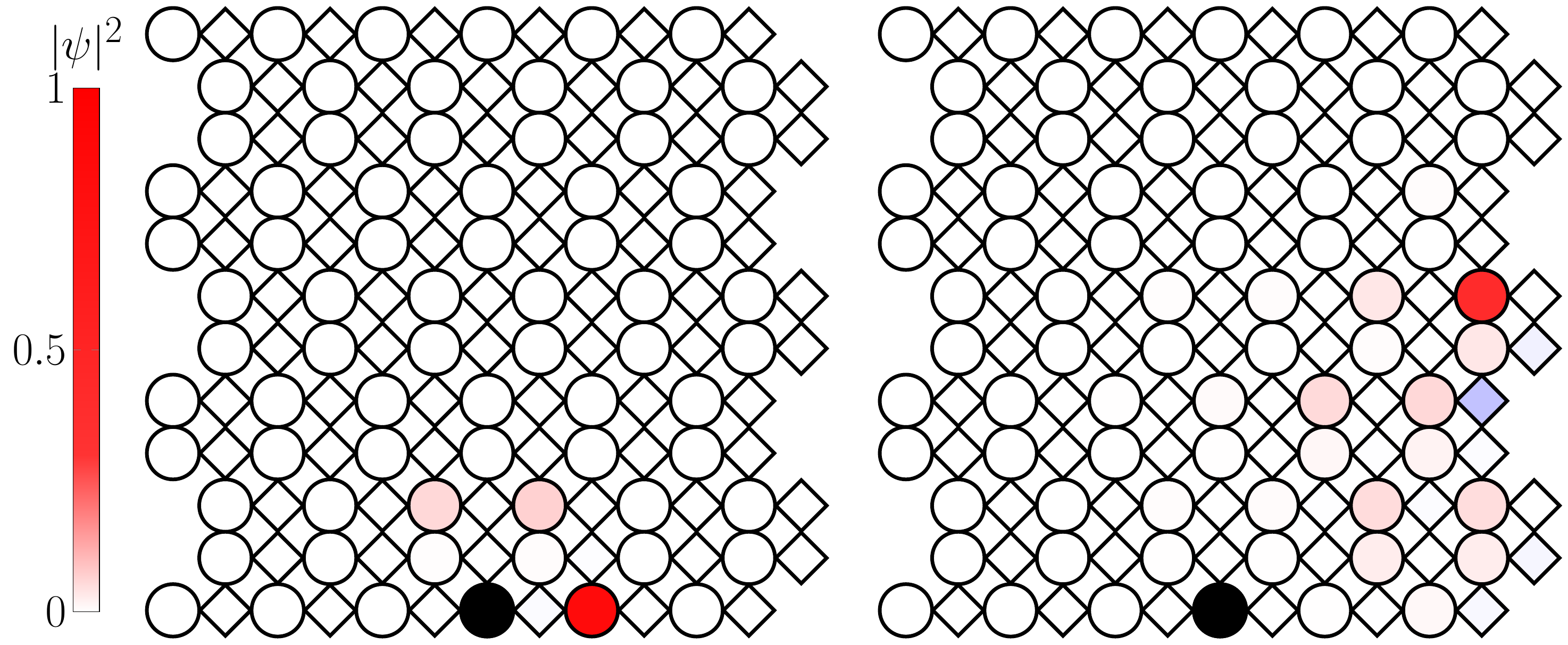}
\end{minipage}
\begin{minipage}[r]{0.5\textwidth}
\includegraphics[scale=0.285]{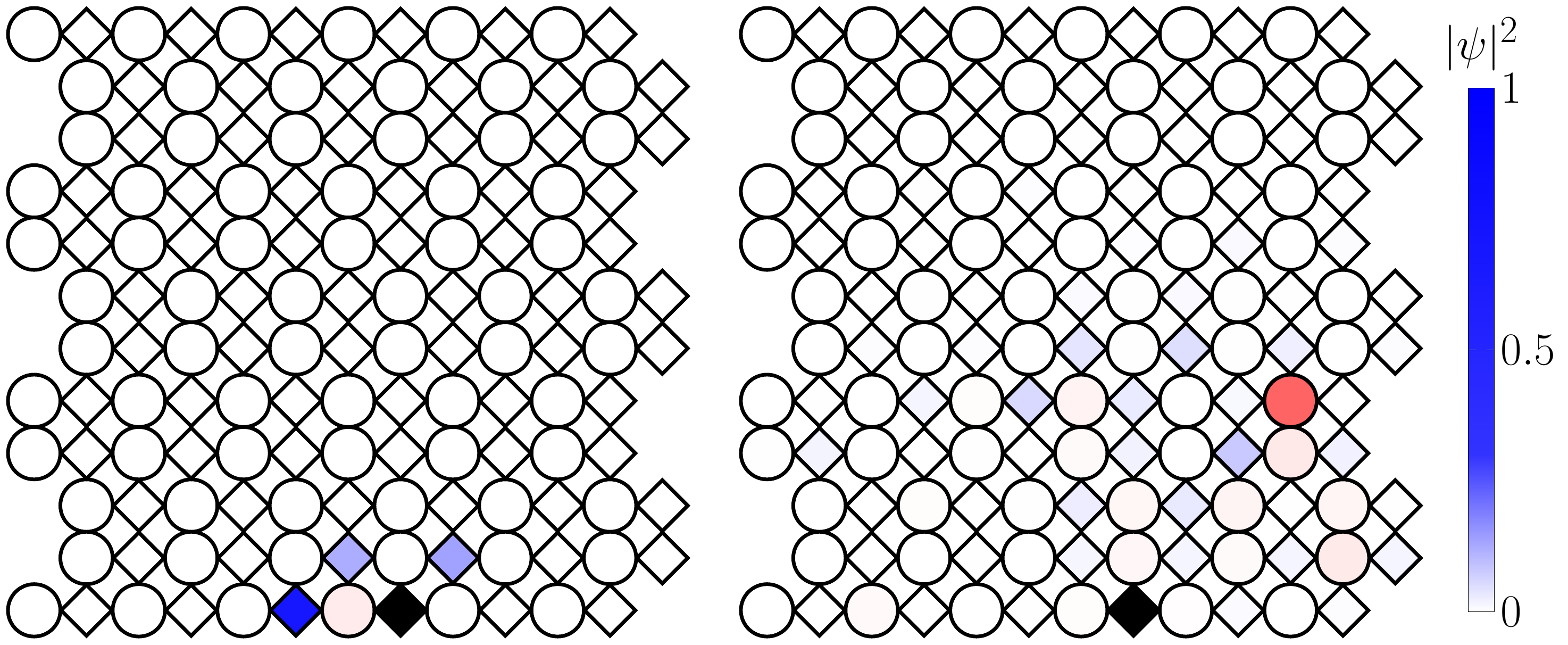}
\end{minipage}
\caption{Edge state propagation on a finite square lattice for fermionic $\mathrm{TRS}^*$ with the parameters from Tab.~\ref{parametersnh}. We use the same edge configuration as in Fig.~\ref{trajec}.  The initial excitation at $t=0$ is marked by a filled black site. The red and blue color indicates the wave function amplitude after one ($t=T$, first and third panel) and five periods ($t=5T$, second and fourth panel) on the two sublattices. After each period, the amplitude at each lattice site is normalized to the total amplitude. }
\label{nhtrans}
\end{figure*}

For this lattice configuration, Eq.~\eqref{eq:model} enforces the patterns of motion shown in Fig.~\ref{trajec} at perfect coupling. Note that we track the patterns of motion for two full cycles because only then the periodicity of the trajectories becomes apparent.  An excitation in the bulk moves in a closed loop, while an excitation starting on an edge site is transported by four sites. Edge states on the sublattice formed by the $A$, $C$ sites (denoted as the red sublattice in the following) move counter-clockwise while their counterparts on the $B$, $D$ sublattice (denoted as the blue sublattice) move clockwise. In waveguide lattices, these trajectories are directly observable.

Tracking the real-space propagation is also a powerful tool to study the properties of the edge states for the non-perfect couplings in Tab.~\ref{parametersnh}. In Fig.~\ref{nhtrans}, we show the real-space propagation of the counterpropagating edge states for fermionic $\mathrm{TRS}^*$ after one ($t=T$) and five ($t=5T$) cycles of the model. Since the parameter values in Tab.~\ref{parametersnh} are close to perfect coupling, we would expect that an initial state at $t=0$, which is prepared on an edge of the red sublattice (blue sublattice), will lead to predominantly counter-clockwise (clockwise) propagation along the lattice perimeter. For fermionic $\mathrm{TRS}^T$, where the edge states have identical imaginary parts, we observe this helical transport. For fermionic $\mathrm{TRS}^*$, however, the two edge states have opposite imaginary part and so the clockwise moving edge state with negative imaginary parts will be suppressed after a few periods. Therefore, we observe counter-clockwise propagation for an initial excitation on the blue sublattice in the long-time limit, and thus chiral, instead of helical, transport.

\begin{figure}
\centering
\includegraphics[width=1\linewidth]{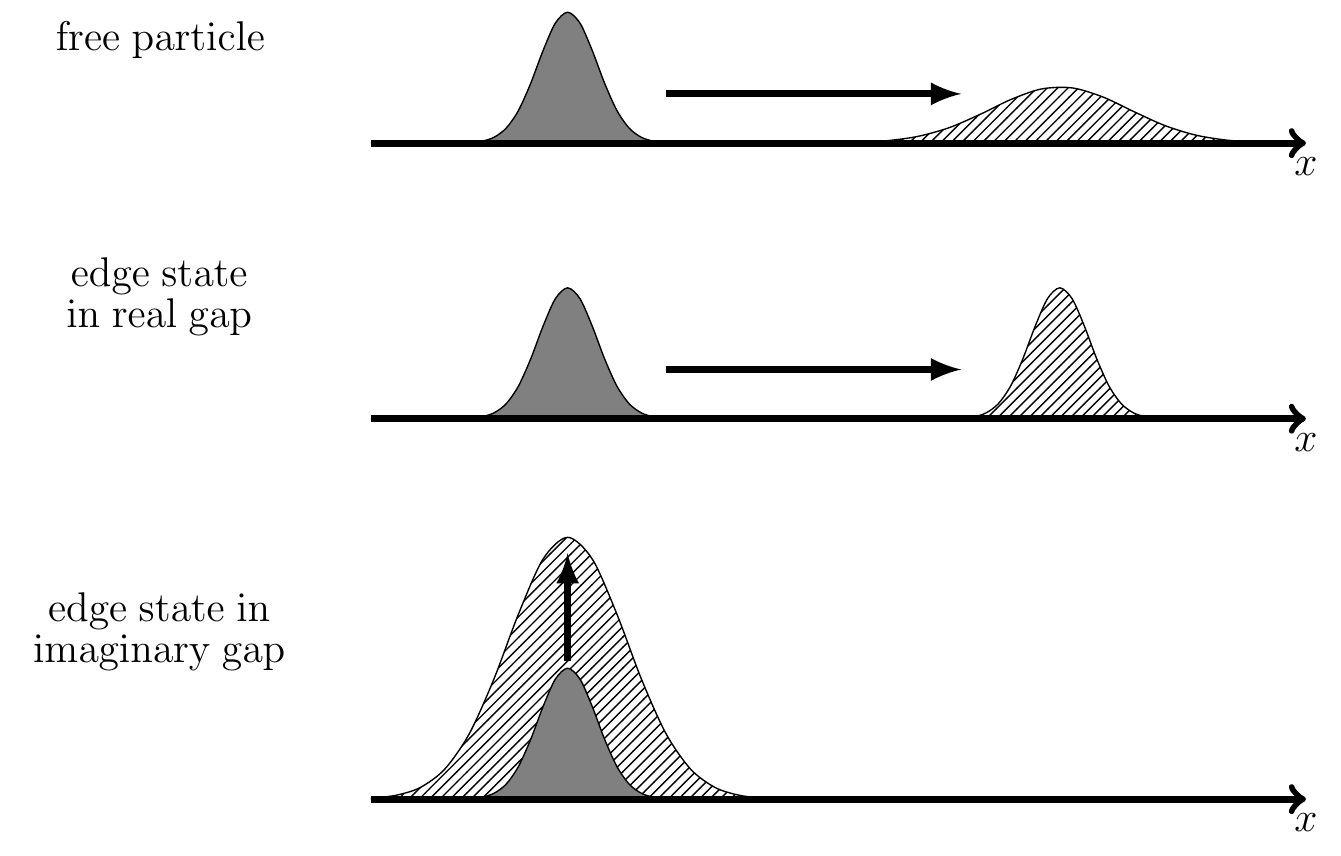}
\caption{Qualitative time evolution of a Gaussian wave packet~\eqref{gaussian} for a free particle with $k_0\ne 0$, an edge state in a real gap with arbitrary $k_0$, and an edge state in an imaginary gap with $k_0=0$. The arrows indicate the time evolution from the initial state (gray) to the final state (hatch pattern). Note that in the bottom panel, the Gaussian does not spread, only its amplitude increases.}
\label{wavepack}
\end{figure}

The edge state in the imaginary gap at $\Im \varepsilon=0$ does not lead to real-space propagation due to the flat real part of its quasienergy dispersion. To illustrate this, we show in Fig.~\ref{wavepack} the stroboscopic time evolution of a one-dimensional Gaussian wave packet
\begin{equation}
\begin{aligned}
\psi_{nT}(x)=\frac{1}{\sqrt{2\pi}} \int_{-\pi}^{\pi}\mathrm{d}k  \;   &\exp\big(\ii k x-\ii  nT\varepsilon(k)\big) 
\\
\times &\exp\big(-k^2/(2\sigma^2)\big)
\end{aligned}
\label{gaussian}
\end{equation}
with momentum width $\sigma$ after multiple driving periods $nT$ for three different quasienergy dispersions: the dispersion of a free particle, an edge state in a real gap, and an edge state in an imaginary gap.

A free particle with $\varepsilon(k) =v (k-k_0)^2$ spreads out during propagation, while an edge state in a real gap with $\varepsilon(k) = v (k-k_0)$ propagates without spreading. The slope $v$ and momentum shift $k_0$ are free parameters. For an edge state in an imaginary gap with $\varepsilon(k) = \mathrm{i} v (k-k_0)$, we get 
\begin{equation}
|\psi_{nT}(x)|^2\simeq \sigma^2 \exp\Big((\sigma vnT)^2-2 k_0 vnT-(\sigma x)^2\Big) \; ,
\label{gaussian_imag}
\end{equation}
if the momentum width $\sigma$ is small relative to the Brillouin zone $[-\pi,\pi)$.
This means that the wave packet is pinned at its initial position and does not spread.  For finite $k_0$, the sign of the slope determines if the amplitude of the wave packet increases or decreases on short time scales. 

For the edge state in Fig.~\ref{imagstates}, the momentum shift $k_0$ is set to zero by bosonic $\mathrm{TRS}^*$. In this case, the amplitude of the wave packet in Eq.~\eqref{gaussian_imag} continuously increases during time evolution.  As long as the imaginary gap stays open, we have $ \exp\Big((\sigma vnT)^2\Big)>1$. The amplification is symmetry-protected.

\section{Conclusion}
\label{sec:conclusion}

The stacked Floquet honeycomb model introduced in the present paper allows for the realization of two counterpropagating edge states in real gaps and a single edge state in an imaginary gap. The four TRS types determine which of the two edge state configurations is possible for a specific parameter set. Switching between the four symmetry types only requires adjustments of the on-site potentials or the sign of the interlayer coupling.
In case of the counterpropagating edge states, the edge transport can be tailored to be either helical or chiral with the imaginary part of the on-site potentials. The edge state in the imaginary gap does not propagate. Instead, it behaves like a localized flat edge state, but with a continuously increasing amplitude.  Our results suggest that the amplification is symmetry-protected by bosonic $\mathrm{TRS}^*$. This essentially follows from the pinned momentum of the edge state. Note that momentum is not conserved in disordered systems. Therefore, the robustness of the symmetry protection with regard to disorder remains an open question for future studies.

Our model provides an experimentally accessible platform for the realization of these edge states in photonic waveguide lattices. 
Since waveguides are gainless, the amplification of the edge state in the imaginary gap would be relative to a uniform loss background in such systems. Inherent amplification is possible in fiber loop setups~\cite{Weidemann311} which also allow for two-dimensional Floquet protocols~\cite{PhysRevA.100.063830,PhysRevLett.121.100502} like our model.

Irrespective of the concrete experimental setup, a fundamental challenge regarding the observation of the edge states in imaginary gaps is the fact that there is always at least one bulk band which has a larger imaginary part than the edge states. Therefore, precise control over the selective excitation of bulk and edge states will prove to be essential in experiments.

For the counterpropagating edge states in the real gaps, this problem can be completely avoided. Since the two edge states are anomalous, non-Hermitian boundary state engineering~\cite{StateEngineering} can be used to tailor the imaginary part of the edge states such that they become dominant relative to the bulk states.

\section*{Authors contribution statement}
All authors specified the scope and the strategy of the paper. The 
calculations were performed by AF and, in part, by BH.  AF and BH wrote 
the manuscript which was edited by all authors.

%\bibliography{referenzen}   % name your BibTeX data base
%\bibliography{referenzen.bib}

\bibliographystyle{spphys}       % APS-like style for physics

\end{document}